%% file: indexCoding_Final_arxiv.tex
\documentclass[10pt,conference]{IEEEtran}
\usepackage{amsmath,amssymb,mathrsfs}
\usepackage{graphicx}
\usepackage{psfrag}
\usepackage{subfigure}
\usepackage{cite}
\usepackage{multirow}
\usepackage{stfloats}

\input{defns}  % default definitions 

\newtheorem{theorem}{Theorem}

\newtheorem{proposition}{Proposition}

\newtheorem{example}{Example}
\newtheorem{corollary}{Corollary}

\title{On the Capacity Region for Index Coding}
\author{
\authorblockN{Fatemeh Arbabjolfaei, Bernd Bandemer, Young-Han Kim,
Eren \c{S}a\c{s}o\u{g}lu, and Lele Wang
}
\authorblockA{Department of Electrical and Computer Engineering\\
University of California, San Diego\\
La Jolla, CA 92093, USA\\
Email: \{farbabjo, bandemer, yhk, esasoglu, lew001\}@ucsd.edu}}

\begin{document}
\maketitle

\begin{abstract}
A new inner bound on the capacity region of the general index coding problem
is established. Unlike most existing bounds that are based on graph
theoretic or algebraic tools, the bound relies on a random coding scheme and
optimal decoding, and has a simple polymatroidal single-letter expression. The
utility of the inner bound is demonstrated by examples that include the capacity
region for all index coding problems with up to five messages (there are 9846
nonisomorphic ones).
\end{abstract}

\vspace{2mm}
\section{Introduction}

\begin{figure}[h!]
\vspace*{2mm}  % FIXME: spacing
\begin{center}
\small
\psfrag{x}[b]{$M_1,\ldots,M_N$}
\psfrag{m1}[b]{$X^n$}
\psfrag{e1}[c]{Encoder}
\psfrag{d1}[c]{Decoder $1$}
\psfrag{d0}[c]{Decoder $2$}
\psfrag{d3}[c]{Decoder $N$}
\psfrag{xh1}[b]{$\Mh_1$}
\psfrag{xh0}[b]{$\Mh_2$}
\psfrag{xh3}[b]{$\Mh_N$}
\psfrag{a1}[b]{$M_{\Ac_1}$}
\psfrag{a2}[b]{$M_{\Ac_2}$}
\psfrag{a3}[b]{$M_{\Ac_N}$}
\includegraphics[scale=0.36]{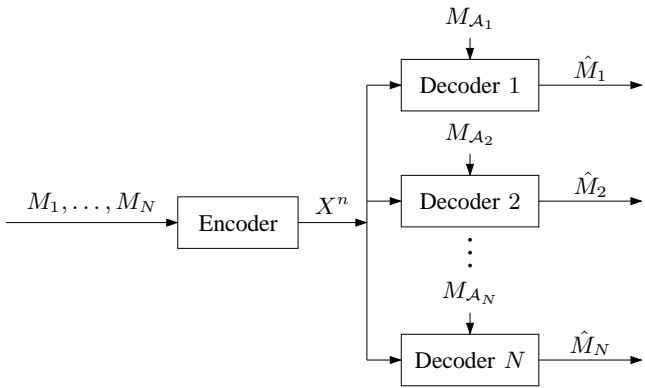}
\end{center}
\caption{The index coding problem.}
\label{fig:index_coding}
\end{figure}

Consider the simple communication problem in Figure~\ref{fig:index_coding},
which is often referred to
as the \emph{index coding} problem.
The sender wishes to communicate $N$ messages $M_j \in [1::2^{nR_j}]$, $j \in
[1::N]$, 
to their respective receivers over a common noiseless link that carries $n$ bits
$X^n$.
Each receiver $j \in [1::N]$ has prior knowledge of $M_{\Ac_j}$, i.e., a subset
$\Ac_j \subseteq [1::N]
\setminus \{j\}$ of the messages.
Based on this side information $M_{\Ac_j}$ and
the received bits $X^n$, receiver $j$ finds the estimate $\Mh_j$ of the message
$M_j$.
A nontrivial tradeoff arises between the rates $R_j$, $j \in [1::N]$, of the
messages since receivers with incompatible knowledge compete
for the shared broadcast medium.

We define a $(2^{nR_1}, \ldots, 2^{nR_N}, n)$ code for index coding by an
encoder $x^n(m_1,\ldots, m_N)$
and $N$ decoders $\mh_j(x^n, m_{\Ac_j})$, $j \in [1::N]$. We assume that the
message tuple $(M_1, \ldots, M_N)$
is uniformly distributed over $[1::2^{nR_1}] \times \cdots \times [1::2^{nR_N}]$, that is, the
messages are uniformly distributed and independent
of each other.
The average probability of error is then defined as $\pen =  \P\{ (\Mh_1,\ldots,
\Mh_N) \ne (M_1, \ldots, M_N)\}$.
A rate tuple $(R_1,\ldots,R_N)$ is said to be achievable if there exists a
sequence of $(2^{nR_1}, \ldots, 2^{nR_N}, n)$  codes such that
$\lim_{n\to\infty} \pen = 0$. The capacity region $\Cr$ of the index coding
problem
is the closure of the set of achievable rate tuples $(R_1,\ldots, R_N)$. (Similarly, one can define the zero-error capacity region, which is
shown in~\cite{Langberg--Effros2011} to be identical to the capacity region.)
The goal is to find the capacity region
and the optimal coding scheme that achieves it.

\begin{figure}[h]
\vspace*{2mm}  % FIXME: spacing
\begin{center}
\subfigure[]{
\small
\psfrag{1}[cb]{1}
\psfrag{2}[r]{2}
\psfrag{3}[l]{3}
\psfrag{4}{4}
\includegraphics[scale=0.45]{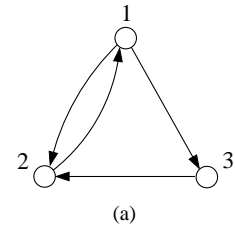}	
}
%\qquad\qquad\qquad
\\[4mm]  % FIXME: spacing
\subfigure[]{
\small
\psfrag{m1}[cr]{$M_1$}
\psfrag{m2}[cr]{$M_2$}
\psfrag{m3}[cr]{$M_3$}
\psfrag{mh1}[cl]{$\Mh_1$}
\psfrag{mh2}[cl]{$\Mh_2$}
\psfrag{mh3}[cl]{$\Mh_3$}
\raisebox{0.1em}{\includegraphics[scale=0.45]{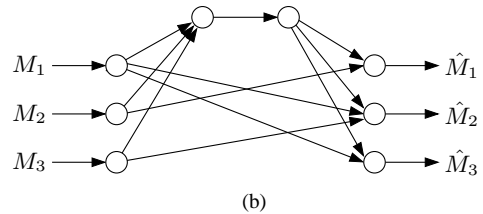}}
}
\end{center}
\caption{(a) Directed graph representation.
(b) The equivalent network coding problem. Here every edge of the graph can
carry up to 1 bit per transmission.}
\label{fig:4-message}
\end{figure}

Note that an index coding problem is fully characterized by the side information
sets $\Ac_j$, $j \in [1::N]$.
As an example, consider the 3-message index coding problem with
$\Ac_1 = \{2\}$, $\Ac_2 = \{1,3\}$, and
$\Ac_3 = \{1\}$.
We represent this problem compactly as
\begin{equation} \label{eq:3-message}
(1|2),\,
(2|1,3),\,
(3|1),
\end{equation}
or as a directed graph (see Figure~\ref{fig:4-message}(a)), where
nodes represent indices of the messages/receivers and edges represent
availability of side information (e.g., the edge $1 \to 2$ means that
side information $M_1$ is available at receiver 2). In general,
an index coding problem $(j | \Ac_j)$, $j \in [1::N]$, can be
represented by a directed graph $\Gc = (\Vc, \Ec)$, where $\Vc =
[1::N]$ and $(j,k) \in \Ec$ iff $j \in \Ac_k$.

Note that the 3-message index coding problem in~\eqref{eq:3-message}
can be represented as an instance of the network coding 
problem~\cite{Ahlswede--Cai--Li--Yeung2000} 
as illustrated in Figure~\ref{fig:4-message}(b).
The same observation can be made for any index coding problem; thus, index
coding is a special case of network
coding.

\medskip

First introduced by Birk and Kol \cite{Birk--Kol1998} in the context of
satellite broadcast communication, the index coding problem has been
studied extensively over the past six years in the theoretical computer science
and network coding communities
with many contributions of combinatorial and algebraic flavors
(see, for example,
\cite{%
Alon--Lubetzky--Stav--Weinstein--Hassidim2008,
Wu--Padhye--Chandra--Padmanabhan--Chou2009, 
Lubetzky--Stav2009,
El-Rouayheb--Sprintson--Georghiades2010,
Bar-Yossef--Birk--Jayram--Kol2011, 
Berliner--Langberg2011,
Blasiak--Kleinberg--Lubetzky2011a, 
Blasiak--Kleinberg--Lubetzky2011b,
Dau--Skachek--Chee2012b,
Effros--El-Rouayheb--Langberg2012,
Haviv--Langberg2012,
Tehrani--Dimakis--Neely2012,
Shanmugam--Dimakis--Langberg2013
}
and the references therein).
Our Shannon-theoretic formulation of the problem closely follows that of
Maleki, Cadambe, and Jafar \cite{Maleki--Cadambe--Jafar2012a},
who established the capacity region for several interesting classes of index
coding problems 
using interference alignment~\cite{Cadambe--Jafar2008}.
Despite all these developments, the capacity region of a general index coding
problem is not known.

Confirming Maslow's axiom~\cite{Maslow1966} ``if all you have is a
hammer, everything looks like a nail,'' we propose a \emph{random
coding} approach, in contrast to more advanced coding schemes of an
algebraic nature.  This approach is more in the spirit of the original
paper by Ahlswede, Cai, Li, and
Yeung~\cite{Ahlswede--Cai--Li--Yeung2000}, where random coding
(binning) was used to establish the network coding theorem.  In
particular, we develop a \emph{composite coding} scheme based on
random coding and establish a corresponding single-letter inner bound
on the capacity region.

Instead of mechanical proofs, this paper focuses on basic intuitions behind our
coding scheme, which we develop gradually from simpler coding schemes---``flat
coding'' in Section~\ref{sec:flat} and ``dual index coding'' in
Section~\ref{sec:dual}. The composite coding scheme is explained in
Section~\ref{sec:composite}. The next section discusses known outer bounds on
the capacity region.

\vspace{1mm}
\section{Outer Bounds}

We first recall the following outer bound on the capacity region (see, for
example, \cite{Blasiak--Kleinberg--Lubetzky2011a}
or \cite{Dougherty--Freiling--Zeger2011} for a similar bound in the context of a
general network coding problem).

\smallskip

\begin{theorem}
\label{thm:outer}
Let $\Bc_j = [1::N] \setminus (\{j\} \cup \Ac_j)$ be the index set of
interfering messages at receiver $j$.
If $(R_1,\ldots, R_N)$ is achievable, then it must satisfy
\[
R_j \le T_{\{j\}\cup \Bc_j} - T_{\Bc_j}, \quad j \in [1::N],
\]
for some $T_\Jc$, $\Jc \subseteq [1::N]$, such that 
\begin{enumerate}
\item $T_\emptyset = 0$, 
\item $T_{[1::N]} = 1$,
\item $T_\Jc \le T_\Kc$ for all $\Jc \subseteq \Kc \subseteq [1::N]$, and
\item $T_{\Jc\cap\Kc} + T_{\Jc\cup\Kc} \le T_\Jc + T_\Kc$ for all $\Jc, \Kc
\subseteq [1::N]$.
\end{enumerate}
\end{theorem}

\smallskip

The upper bound is established by using Fano's
inequality and setting $T_\Jc=(1/n)\,H(X^n\cond M_{\Jc^c})$.  Properties
1--4 are due to the submodularity of entropy.

Recent results by Sun and Jafar~\cite{Sun--Jafar2013} indicate that this outer
bound is not tight in general. Nevertheless, a 
relaxed version of the bound is sometimes useful.

\smallskip

\begin{corollary}
\label{cor:cycle}
If $(R_1,\ldots,R_N)$ is achievable for an index coding problem represented by
the directed graph
$\Gc$, then it must satisfy
\[
\sum_{j \in \Jc} R_j \le 1
\]
for all $\Jc \subseteq [1::N]$ for which the subgraph of $\Gc$ over $\Jc$ does
not contain a directed cycle.
\end{corollary}
\smallskip
The following example, due to \cite{Tehrani--Dimakis--Neely2012,
Maleki--Cadambe--Jafar2012a},
illustrates that the two outer bounds do not coincide in general.

\smallskip

\begin{example}
\label{eg:5-node}
Consider the symmetric five-message index coding problem $(j \cond j-1,j+1)$, $j
\in [1::5]$, namely,
\[
(1|5,2),\, (2|1,3),\, (3|2,4),\, (4|3,5),\, (5|4,1).
\]
The corresponding graph representation is depicted in Figure~\ref{fig:5-node}.
\begin{figure}[t]
\vspace*{2mm}  % FIXME: spacing
\begin{center}
\small
\psfrag{1}[cb]{1}
\psfrag{2}[lb]{2}
\psfrag{3}[lt]{3}
\psfrag{4}[rt]{4}
\psfrag{5}[rb]{5}
\includegraphics[scale=0.45]{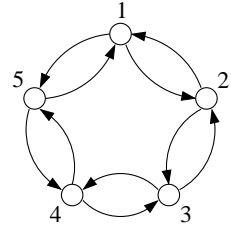}	
\end{center}
\caption{A graph representation of the 5-message index coding problem.}
\label{fig:5-node}
\end{figure}
Applying Corollary~\ref{cor:cycle}, we obtain 
\begin{equation} \label{eq:5-node-ineq1}
\begin{split}
R_1 + R_3 &\le 1,\quad
R_2 + R_4 \le 1,\quad
R_3 + R_5 \le 1,\\
R_4 + R_1 &\le 1,\quad
R_5 + R_2 \le 1.
\end{split}
\end{equation}
In comparison, Theorem~\ref{thm:outer} leads to the inequality
\begin{equation} \label{eq:5-node-ineq2}
R_1 + R_2 + R_3 + R_4 + R_5 \le 2,
\end{equation}
in addition to the five inequalities above. 
As we discuss in Section~\ref{sec:composite},
\eqref{eq:5-node-ineq1}~and~\eqref{eq:5-node-ineq2}
characterize 
the capacity region of this index coding problem.
\end{example}

%\vspace{1mm}
\section{Flat Coding} \label{sec:flat}

Consider the following simple random coding scheme. 
For each $(m_1,\ldots, m_N) \in [1::2^{nR_1}] \times \cdots \times
[1::2^{nR_N}]$,
generate a codeword $x^n(m_1,\ldots,m_N)$ randomly and independently as a
Bern($1/2$) sequence.
To communicate $(m_1,\ldots, m_N)$, the sender transmits $x^n =
x^n(m_1,\ldots,m_N)$.
Receiver $j$ uses simultaneous nonunique
decoding~\cite{Chong--Motani--Garg--El-Gamal2008} and finds the unique $\mh_j
\in [1::2^{nR_j}]$
such that $x^n(\mh_j, m_{\Ac_j}, m_{\Bc_j})$ is jointly typical with
(i.e., identical to) 
the received sequence $x^n$ for some $m_{\Bc_j}$,
where $\Bc_j = [1::N] \setminus (\{j\} \cup \Ac_j)$.
Since the codebook generation is ``flat'' (compared with ``layered''
superposition coding), simultaneous nonunique decoding is essentially
identical to uniquely decoding $(\mh_j, \mh_{\Bc_j})$ and then
discarding the unnecessary part $\mh_{\Bc_j}$.  This flat coding
scheme yields the following inner bound.

\smallskip

\begin{proposition}
\label{prop:flat}
A rate tuple $(R_1,\ldots,R_N)$ is achievable for the index coding
problem $(j | \Ac_j)$, $j \in [1::N]$, if
\[
R_j + \sum_{k \in \Bc_j} R_k < 1,\quad j \in [1::N].
\]
\end{proposition}

\smallskip

As an example, consider the 3-message problem in~\eqref{eq:3-message}. Under
flat coding,
receiver~1 finds the unique $\mh_1$ such that $x^n(\mh_1, m_2, m_3) = x^n$ for
some $m_3 \in [1::2^{nR_3}]$ and the given side information $m_2$. By the
packing
lemma~\cite[Sec.~3.4]{El-Gamal--Kim2011},
it can be readily shown that the probability of decoding error for receiver~1
tends to zero as
$n \to \infty$ if 
\begin{equation} \label{eq:3-node-ineq1}
R_1 + R_3 < 1.
\end{equation}
Similarly, we obtain
$R_2 < 1$ (an inactive bound) and 
\begin{equation} \label{eq:3-node-ineq2}
R_2 + R_3 < 1.
\end{equation}
By comparing with Theorem~\ref{thm:outer} (or
Corollary~\ref{cor:cycle}), it can be easily checked that the rate
region characterized by~\eqref{eq:3-node-ineq1} and~\eqref{eq:3-node-ineq2} is
indeed the capacity region.

It can be easily verified that for all index coding problems with 1,
2, and 3 messages---there are 1, 3, and 16 nonisomorphic
problems~\cite{OEIS--A000273}---this flat coding scheme
(or more generally, time sharing of flat coding over different subsets of
messages)
achieves the capacity region.
Among the 218 four-message index coding problems,
time sharing of flat coding over subsets of messages achieves the
capacity region for all but three. The following is one of the three
exceptions.

\smallskip

\begin{example} \label{eg:4-node}
Consider the 4-message index coding problem
\begin{equation*} 
%\label{eq:4-message}
(1|4),\,
(2|3,4),\,
(3|1,2),\,
(4|2,3).
\end{equation*}
On the one hand, flat coding yields an inner bound
on the capacity region that consists of the rate quadruples $(R_1,R_2,R_3,R_4)$
such that
\begin{equation*}
\label{eq:4-node-inner}
\begin{split}
R_1 + R_2 + R_3 &< 1,\\
R_1 + R_4 &< 1,\\
R_3 + R_4 &< 1.
\end{split}
\end{equation*}
It can be verified that this inner bound cannot be improved upon by time sharing
over subsets.
On the other hand, Theorem~\ref{thm:outer} (or Corollary~\ref{cor:cycle})
yields an outer bound that consists of
the rate quadruples $(R_1,R_2,R_3,R_4)$ such that
\begin{equation}
\label{eq:4-node-outer}
\begin{split}
R_1 + R_2 &\le 1,\qquad
R_1 + R_3 \le 1,\\
R_1 + R_4 &\le 1,\qquad
R_3 + R_4 \le 1.
\end{split}
\end{equation}
We will see in Section~\ref{sec:composite} that this outer bound is
tight.
\end{example}

\smallskip

While flat coding is suboptimal in general, 
the analysis (i.e., the proof of Proposition~\ref{prop:flat}) is trivial
and does not rely on any graph theoretic machinery.
This observation will be crucial when we generalize the coding scheme
subsequently.

\section{Dual Index Coding} \label{sec:dual}

\begin{figure}[h!]
\vspace*{2mm}  % FIXME: spacing
\begin{center}
\small
\psfrag{x}[b]{$M_1,\ldots,M_N$}
\psfrag{m1}[bc]{$M_1$}
\psfrag{m2}[bc]{$M_\Jc$}
\psfrag{m3}[bc]{$M_1,\ldots,M_N$}
\psfrag{mh}[bc]{$\Mh_1,\ldots,\Mh_N$}
\psfrag{s1}[bc]{$W_1$}
\psfrag{s2}[bc]{$W_\Jc$}
\psfrag{s3}[bc]{$W_{[1::N]}$}
\psfrag{e1}[c]{Encoder $1$}
\psfrag{e2}[c]{Encoder $\Jc$}
\psfrag{e3}[c]{Encoder $[1::N]$}
\psfrag{d1}[c]{Decoder}
\psfrag{d0}[c]{Decoder $2$}
\psfrag{d3}[c]{Decoder $N$}
\psfrag{xh1}[b]{$\Mh_1$}
\psfrag{xh0}[b]{$\Mh_2$}
\psfrag{xh3}[b]{$\Mh_N$}
\psfrag{a1}[b]{$M_{\Ac_1}$}
\psfrag{a2}[b]{$M_{\Ac_2}$}
\psfrag{a3}[b]{$M_{\Ac_N}$}
\includegraphics[scale=0.36]{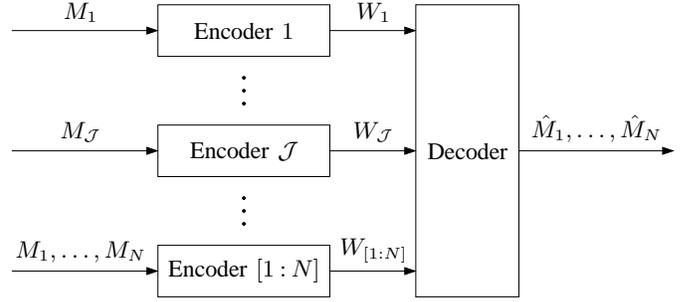}
\\ \vspace*{-2mm}  %FIXME: spacing
\end{center}
\caption{The dual index coding problem.}
\label{fig:dual-index}
\end{figure}

Before we move on to a more powerful random coding scheme, we
introduce a communication problem (depicted in
Figure~\ref{fig:dual-index}) that is, in some sense, dual to the index
coding problem.  Here a set of $(2^N-1)$ senders wish to communicate a
message tuple $(M_1,\ldots,M_N)$ to a common receiver through a
noiseless channel, each encoding a subtuple $M_\Jc$ into a separate
index $W_\Jc \in [1::2^{n S_\Jc}]$ for all nonempty $\Jc
\subseteq [1::N]$.  What is the capacity region (as a function of the
rates $S_\Jc$)?

This problem is a special case of the general multiple access
channel (MAC) with correlated messages studied by Han~\cite{Han1979}.
For the general MAC, superposition coding achieves the capacity region
that is characterized by independent auxiliary random variables $U_1,
\ldots, U_N$, each corresponding to a message.  However, for the dual
index coding problem, we can characterize the capacity region
explicitly.

\smallskip

\begin{proposition}
\label{prop:dual}
The capacity region of the dual index coding problem is
the set of rate tuples $(R_1,\ldots, R_N)$ that satisfy
\begin{equation} \label{eq:dual-index}
\sum_{j \in \Jc}R_j \le \sum_{\Jc' \subseteq [1::N] \suchthat \Jc' \cap \Jc \ne
\emptyset} S_{\Jc'}
\end{equation}
for all $\Jc \subseteq [1::N]$.
\end{proposition}

\smallskip

What is perhaps more important than this explicit characterization of the
capacity region
is the fact that it can be achieved by flat coding, which we will utilize later.

\begin{figure}[h!]
\vspace{3mm}  % FIXME: spacing
\begin{center}
\small
\psfrag{x}[b]{$M_1,\ldots,M_N$}
\psfrag{m1}[bc]{$M_1,M_2$}
\psfrag{m2}[bc]{$M_1,M_3$}
\psfrag{m3}[bc]{$M_1,M_2,M_3$}
\psfrag{mh}[bc]{$\Mh_1,\Mh_2,\Mh_3$}
\psfrag{s1}[bc]{1}
\psfrag{s2}[bc]{2}
\psfrag{s3}[bc]{2}
\psfrag{e1}[c]{Encoder $\{1,2\}$}
\psfrag{e2}[c]{Encoder $\{1,3\}$}
\psfrag{e3}[c]{Encoder $\{1,2,3\}$}
\psfrag{d1}[c]{Decoder}
\psfrag{d0}[c]{Decoder $2$}
\psfrag{d3}[c]{Decoder $N$}
\psfrag{xh1}[b]{$\Mh_1$}
\psfrag{xh0}[b]{$\Mh_2$}
\psfrag{xh3}[b]{$\Mh_N$}
\psfrag{a1}[b]{$M_{\Ac_1}$}
\psfrag{a2}[b]{$M_{\Ac_2}$}
\psfrag{a3}[b]{$M_{\Ac_N}$}
\includegraphics[scale=0.36]{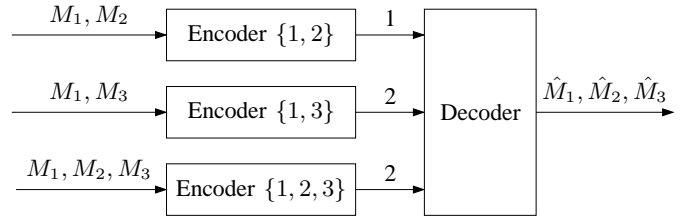}
\\ \vspace*{-3mm}  %FIXME: spacing
\end{center}
\caption{An instance of dual index coding.  %The rates of encoders $\{1\}$, $\{2\}$, $\{3\}$, and $\{2,3\}$ are zero.
}
\label{fig:3-sender}
\end{figure}

As an example, consider the three-message three-sender dual index coding problem
in Figure~\ref{fig:3-sender},
where $S_{1,2} = 1$, $S_{1,3} = S_{1,2,3} = 2$, and $S_{1}=S_{2}=S_{3}=S_{2,3}=0$.
By~\eqref{eq:dual-index}, the capacity region is the set of
rate triples $(R_1,R_2,R_3)$ such that
\[
R_1 + R_2 + R_3 \le 5,\qquad
R_2 \le 3,\qquad
R_3 \le 4.
\]
This can be achieved via flat coding of $(M_1, M_2)$, $(M_1,M_3)$, and
$(M_1,M_2,M_3)$, respectively,
and simultaneous decoding at the receiver.

\section{Composite coding} \label{sec:composite}

Equipped with the results in the previous two subsections, we now
introduce a layered random coding scheme, which we refer to as
\emph{composite coding}. This is best described by an example. 

\begin{figure*}[!b]
\begin{center}
\small
\bigskip
\psfrag{x}[b]{$X^n$}
\psfrag{e}[c]{Encoder}
\psfrag{d1}[c]{Decoder $1$}
\psfrag{d0}[c]{Decoder $2$}
\psfrag{d3}[c]{Decoder $N$}
\psfrag{xh1}[b]{$\Mh_1$}
\psfrag{xh0}[b]{$\Mh_2$}
\psfrag{xh3}[b]{$\Mh_N$}
\psfrag{a1}[b]{$M_{\Ac_1}$}
\psfrag{a2}[b]{$M_{\Ac_2}$}
\psfrag{a3}[b]{$M_{\Ac_N}$}
\psfrag{m1}[bc]{$M_1$}
\psfrag{m2}[bc]{$M_\Jc$}
\psfrag{m3}[bc]{$M_1,\ldots,M_N$}
\psfrag{mh}[bc]{$\Mh_1,\ldots,\Mh_N$}
\psfrag{wh1}[bc]{$\Wh_1, \Wh_2, \ldots, \Wh_{[1::N]}$}
\psfrag{wh0}[bc]{$\Wh_1, \Wh_2, \ldots, \Wh_{[1::N]}$}
\psfrag{wh3}[bc]{$\Wh_1, \Wh_2, \ldots, \Wh_{[1::N]}$}
\psfrag{s1}[bc]{$W_1$}
\psfrag{s2}[bc]{$W_\Jc$}
\psfrag{s3}[bc]{$W_{[1::N]}$}
\psfrag{e1}[c]{Encoder $1$}
\psfrag{e2}[c]{Encoder $\Jc$}
\psfrag{e3}[c]{Encoder $[1::N]$}
\includegraphics[scale=0.36]{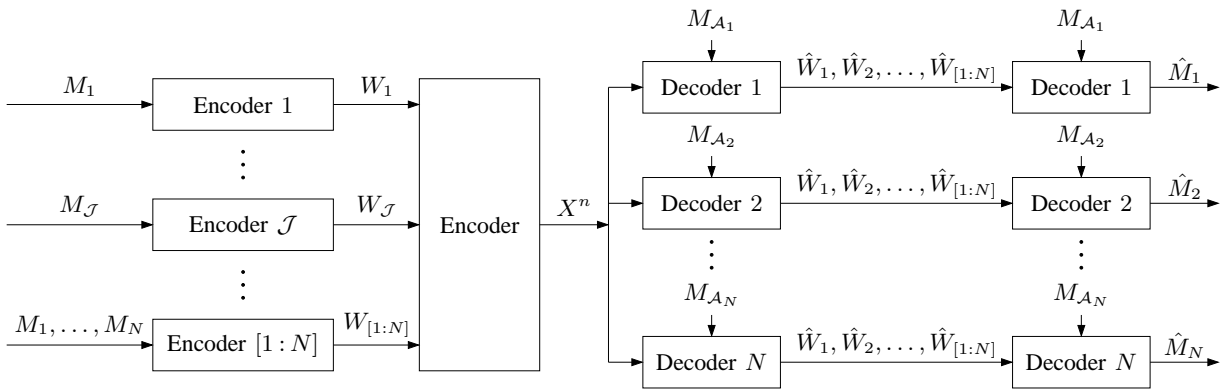}
\end{center}
\caption{Composite coding scheme.}
\label{fig:composite}
\end{figure*}

Consider again the 5-message problem
in Example~\ref{eg:5-node}.
In the first step of composite coding, the sender encodes $(M_1,M_2)$ into an
index $W_{1,2}$ at rate $S_{1,2}$ using random coding, 
and similarly encodes $(M_2,M_3)$, $(M_3,M_4)$, $(M_4,M_5)$, and $(M_1,M_5)$,
respectively, into indices $W_{2,3}$,
$W_{3,4}$, $W_{4,5}$, and $W_{1,5}$. Equivalently, the sender
is decomposed into~5 ``virtual'' senders, each encoding one of the above pairs
of messages (as in the dual index coding problem).
In the second step, the sender uses flat coding to
encode the ``composite'' indices $W_{1,2}, W_{2,3}, W_{3,4}, W_{4,5}, W_{1,5}$.
As with encoding, decoding also takes two steps.
Each receiver first recovers all composite indices, and then recovers the
desired message from the composite indices.
For example, receiver~1 recovers $W_{1,2}, W_{1,5}$ (along with
other composite indices).  Since receiver~1 has side information
$(M_2, M_5)$, it can recover $M_1$ from $(W_{1,2}, W_{1,5})$ if $R_1\le S_{1,2} 
+ S_{1,5}$. Following similar steps for other receivers and
incorporating the flat coding rate condition, it can be easily verified that
a rate quintuple $(R_1,R_2,R_3,R_4,R_5)$ is achievable if
\begin{align*}
R_1 &< S_{1,2} + S_{1,5},\\
R_2 &< S_{1,2} + S_{2,3},\\
R_3 &< S_{2,3} + S_{3,4},\\
R_4 &< S_{3,4} + S_{4,5},\\
R_5 &< S_{1,5} + S_{4,5}
\end{align*}
for some $(S_{1,2}, S_{2,3}, S_{3,4}, S_{4,5}, S_{1,5})$
satisfying
$S_{1,2} + S_{2,3} + S_{3,4} + S_{4,5} + S_{1,5} \le 1$.
Fourier--Motzkin elimination~\cite[Appendix~D]{El-Gamal--Kim2011}
of the composite index rates yields the inequalities
\eqref{eq:5-node-ineq1}~and~\eqref{eq:5-node-ineq2} that define the outer bound, thus establishing the capacity region.

Now consider the four-message problem in
Example~\ref{eg:4-node}. In this case,
we only use the composite indices $W_{1,4}$ and $W_{1,2,3,4}$
with rates
$S_{1,4}$ and $S_{1,2,3,4}$, respectively, and set the
rates of the remaining indices to zero.  It can be easily
verified from Proposition~\ref{prop:dual} that receiver~1 can recover $M_1$ if
$R_1 < S_{1,4}$; receiver~2 can recover $M_2$ (and $M_1$) if
$R_1 + R_2 < S_{1,4} + S_{1,2,3,4}$ and $R_2 < S_{1,2,3,4}$; 
receiver~3 can recover $M_3$ (and $M_4$) if
$R_3 + R_4 < S_{1,4} + S_{1,2,3,4}$ and $R_3 < S_{1,2,3,4}$; and
receiver~4 can recover $M_4$ (and $M_1$) if
$R_1 + R_4 < S_{1,4} + S_{1,2,3,4}$. Adding the constraint
$S_{1,4} + S_{1,2,3,4} \le 1$ and eliminating $S_{1,4}$ and
$S_{1,2,3,4}$, we obtain the inequalities~\eqref{eq:4-node-outer}
that define the outer bound, thus establishing the capacity region.

In general, we can utilize $(2^N - 1)$ virtual senders to encode $N$
messages. Moreover, the receivers can employ simultaneous nonunique
decoding in the second step (or equivalently, ignore some
of the composite indices, as in the examples above).
This coding scheme is illustrated in Figure~\ref{fig:composite}.
It easily follows from the arguments above that if we allow decoder $j$
to decode a subset $\Kc_j$ of the messages, then the rates of the
composite messages need to belong to the polymatroidal rate region
$\Rr( \Kc_j \cond \Ac_j )$ defined by
\begin{equation} \label{eq:polymatroid}
\sum_{j\in\Jc} R_j < \sum_{\Jc' \subseteq \Kc_j \cup \Ac_j \suchthat \Jc' \cap
\Jc \ne \emptyset} S_{\Jc'}
\end{equation}
for all $\Jc \subseteq \Kc_j \setminus \Ac_j$. Note that this is the capacity region of the dual index coding problem
(Proposition~\ref{prop:dual}) with message set $\Kc_j$ and side
information $\Ac_j$.  Taking the union over all choices of
decoding sets $\Kc_j$ yields the following inner bound, which is the
main result of the paper.
\smallskip

\begin{theorem}[Composite-coding inner bound]
\label{thm:composite}
A rate tuple $(R_1,\ldots, R_N)$ is achievable for
the index coding problem $(j | \Ac_j)$, $j \in [1::N]$,
if 
\begin{equation} \label{eq:composite}
(R_1,\ldots, R_N) \in \bigcap_{j \in [1::N]} \;\;\bigcup_{
\Kc_j \subseteq [1:N] \suchthat j \in \Kc_j
} \Rr( \Kc_j \cond \Ac_j)
\end{equation}
for some $(S_\Jc \suchthat \Jc \subseteq [1::N])$
such that $\sum_{\Jc \suchthat \Jc
\not \subseteq \Ac_j} S_\Jc \le 1$
for all $j \in [1::N]$.
\end{theorem}

\smallskip

At first glance, composite coding seems to be time sharing of flat coding over
all subsets of $[1::N]$. However, it employs the optimal decoding rule that
utilizes all composite indices (subsets) that are relevant to the desired
message. As such, the corresponding rate region has a very similar form as the
optimal rate region for interference networks with random
coding~\cite{Bandemer--El-Gamal--Kim2012}.

Using the polco tool for polyhedral computations~\cite{Terzer2009}, we have
computed the composite-coding inner bound 
and the outer bound in Theorem~\ref{thm:outer}
for all 9608 nonisomorphic five-message index coding
problems~\cite{OEIS--A000273}. In all cases, inner and outer bounds agree,
establishing the capacity region. 

To further demonstrate the utility of composite coding, we revisit the following
example
in~\cite{Maleki--Cadambe--Jafar2012a}.

\smallskip

\begin{example}
Consider the $N$-message \emph{symmetric} index coding problem
\[
(j \cond j-U, j-U+1, \ldots, j-1, j+1, \ldots, j+D)
\]
for $j \in [1::N]$, where all message indices are understood modulo $N$.
For instance, the 5-message problem in Example~\ref{eg:5-node} is a special case
of this problem with $N = 5$ and $D = U = 1$.
We assume without loss of generality that $0 \le U \le D \le N-U-1$.
Let $S_\Jc = 1/(N-(D-U))$ if $\Jc$ is of the form $[k::k+U]$, and let $S_\Jc = 0$ otherwise.
Since receiver $j \in [1::N]$ has $M_{j+1}, \ldots, M_{j+D}$ as side
information, 
it already knows the $D-U$ composite indices $W_{[j+1::j+1+U]}, \ldots,
W_{[j+D-U::j+D]}$. Thus, there are only $N-(D-U)$ composite indices that need
to be recovered from $x^n$,
which is feasible since 
$\sum_{\Jc \suchthat \Jc \not \subseteq \Ac_j} S_\Jc = 1$.
Now receiver $j$ can recover $M_j$ from the composite indices
$W_{[j-U::j]}, \ldots, W_{[j::j+U]}$, provided that
\[
R_j < S_{[j-U::j]} + \cdots + S_{[j:j+U]}.
\]
Hence, the symmetric rate of $(U+1)/(N-D+U)$ is achievable. In
\cite{Maleki--Cadambe--Jafar2012a} it is shown that this symmetric rate is in
fact optimal, which can be also verified directly
by the outer bound in Theorem~\ref{thm:outer}.
For $N = 6, U = 1$, and $D = 2$, that is,
\begin{align*}
&(1|2,3,6), (2|1,3,4), (3|2,4,5),\\
&(4|3,5,6), (5|4,6,1), (6|5,1,2),
\end{align*}%
the symmetric rate of $2/5$ is optimal. In fact, simplifying
Theorems~\ref{thm:outer} and~\ref{thm:composite} yields the capacity region that
consists of the rate sextuples $(R_1,\ldots,R_6)$
such that
\begin{align*}
R_j + R_{j+2} &\le 1,\quad j \in [1::6],\\
R_j + R_{j+3} &\le 1,\quad j \in [1::6],\\
R_j + R_{j+1} + R_{j+2} + R_{j+3} + R_{j+4} &\le 2,\quad j \in [1::6].
\end{align*}
In particular, this region is achievable by using composite indices $W_1$,
$W_2$, $W_3$, $W_4$, 
$W_5$, $W_6$,
$W_{1,2}$, $W_{2,3}$, $W_{3,4}$, $W_{4,5}$, $W_{5,6}$, and $W_{1,6}$.
\end{example}

\section{Concluding Remarks}

Based on a first principle in Shannon's random coding, this paper has
established the composite-coding inner bound on the general index
coding problem.  This inner bound is simple, easy to compute, yet is
tight for all index coding problems of up to five
messages as well as many existing examples.  In a sense, random coding
is a ``jackknife'' rather than a ``hammer.''

The polymatroidal structure of the composite-coding inner bound and the
submodularity of the 
outer bound suggest a deeper connection rooted in matroid
theory~\cite{Oxley2006, Dougherty--Freiling--Zeger2011}. In addition to
evaluating the inner and outer bounds for more examples
(there are 1540944 nonisomorphic six-message index coding problems), future
studies will focus on analyzing the algebraic structures of these bounds to
investigate what lies in the path to establishing the capacity region of a
general index coding problem.

%------------------------------------------------------------------------------

\newcommand{\BIBdecl}{\addtolength{\itemsep}{0.38mm}}   % FIXME: change this to fill the page
\bibliographystyle{IEEEtran}
\bibliography{nit}

\end{document}

%% file: defns.tex
\usepackage{xspace}
\usepackage{bbm}
\input{mathlig}

\newcommand{\muspace}{\mspace{1mu}}

\DeclareRobustCommand{\scond}{\mathchoice{\muspace\vert\muspace}{\vert}{\vert}{\vert}}
\mathlig{|}{\scond}
\newcommand{\cond}{\mathchoice{\,\vert\,}{\mspace{2mu}\vert\mspace{2mu}}{\vert}{\vert}}

\DeclareRobustCommand{\discint}{\mathchoice{\mspace{-1.5mu}:\mspace{-1.5mu}}{\mspace{-1.5mu}:\mspace{-1.5mu}}{:}{:}}
\mathlig{::}{\discint}
\newcommand{\suchthat}{\mathchoice{\colon}{\colon}{:\mspace{1mu}}{:}}

\newcommand{\Ac}{\mathcal{A}}
\newcommand{\Bc}{\mathcal{B}}

\newcommand{\Ec}{\mathcal{E}}

\newcommand{\Gc}{\mathcal{G}}

\newcommand{\Jc}{\mathcal{J}}
\newcommand{\Kc}{\mathcal{K}}

\newcommand{\Vc}{\mathcal{V}}

\newcommand{\Cr}{\mathscr{C}}

\newcommand{\Rr}{\mathscr{R}}

\newcommand{\pen}{{P_e^{(n)}}}

\newcommand{\Mh}{{\hat{M}}}

\newcommand{\Wh}{{\hat{W}}}

\newcommand{\mh}{{\hat{m}}}

\let\P\relax
\DeclareMathOperator\P{\textsf{P}}

\def\textiid{i.i.d.\@\xspace}
\newcommand\iid{\ifmmode\text{ i.i.d. } \else \textiid \fi}

\def\mathllap{\mathpalette\mathllapinternal}
\def\mathllapinternal#1#2{%
  \llap{$\mathsurround=0pt#1{#2}$}}

\def\clap#1{\hbox to 0pt{\hss#1\hss}}
\def\mathclap{\mathpalette\mathclapinternal}
\def\mathclapinternal#1#2{%
  \clap{$\mathsurround=0pt#1{#2}$}}

\let\oldstackrel\stackrel
\renewcommand{\stackrel}[2]{\oldstackrel{\mathclap{#1}}{#2}}

\renewcommand{\hbar}{h\mathllap{\overline{\vphantom{h}\hphantom{\rule{4.6pt}{0pt}}}\mspace{0.77mu}}}

\catcode`~=11 % make LaTeX treat tilde (~) like a normal character
\newcommand{\urltilde}{\kern -.06em\lower -.06em\hbox{~}\kern .02em}
\catcode`~=13 % revert back to treating tilde (~) as an active character

%% file: mathlig.tex
\catcode`@=11
\chardef\mathlig@atcode\count255

\def\actively#1#2{\begingroup\uccode`\~=`#2\relax\uppercase{\endgroup#1~}}
\def\mathlig@gobble{\afterassignment\mathlig@next@cmd\let\mathlig@next= }
\def\mathlig@delim{\mathlig@delim}
\def\mathlig@defcs#1{\expandafter\def\csname#1\endcsname}
\def\mathlig@let@cs#1#2{\expandafter\let\expandafter#1\csname#2\endcsname}
\def\mathlig@appendcs#1#2{\expandafter\edef\csname#1\endcsname{\csname#1\endcsname#2}}

\def\mathlig#1#2{\mathlig@checklig#1\mathlig@end\mathlig@defcs{mathlig@back@#1}{#2}\ignorespaces}

\def\mathlig@checklig#1#2\mathlig@end{%
 \expandafter\ifx\csname mathlig@forw@#1\endcsname\relax
 \expandafter\mathchardef\csname mathlig@back@#1\endcsname=\mathcode`#1%
 \mathcode`#1"8000\actively\def#1{\csname mathlig@look@#1\endcsname}%
 \mathlig@dolig#1\mathlig@delim
\fi
\mathlig@checksuffix#1#2\mathlig@end
}

\def\mathlig@checksuffix#1#2\mathlig@end{%
\ifx\mathlig@delim#2\mathlig@delim\relax\else\mathlig@checksuffix@{#1}#2\mathlig@end\fi
}
\def\mathlig@checksuffix@#1#2#3\mathlig@end{%
\expandafter\ifx\csname mathlig@forw@#1#2\endcsname\relax\mathlig@dosuffix{#1}{#2}\fi
\mathlig@checksuffix{#1#2}#3\mathlig@end
}

\def\mathlig@dosuffix#1#2{%
\mathlig@appendcs{mathlig@toks@#1}{#2}%
\mathlig@dolig{#1}{#2}\mathlig@delim
}

\def\mathlig@dolig#1#2\mathlig@delim{%
 \mathlig@defcs{mathlig@look@#1#2}{%
 \mathlig@let@cs\mathlig@next{mathlig@forw@#1#2}\futurelet\mathlig@next@tok\mathlig@next}%
 \mathlig@defcs{mathlig@forw@#1#2}{%
  \mathlig@let@cs\mathlig@next{mathlig@back@#1#2}%
  \mathlig@let@cs\checker{mathlig@chck@#1#2}%
  \mathlig@let@cs\mathligtoks{mathlig@toks@#1#2}%
  \expandafter\ifx\expandafter\mathlig@delim\mathligtoks\mathlig@delim\relax\else
  \expandafter\checker\mathligtoks\mathlig@delim\fi
  \mathlig@next
 }%
 \mathlig@defcs{mathlig@toks@#1#2}{}%
 \mathlig@defcs{mathlig@chck@#1#2}##1##2\mathlig@delim{%
  \ifx\mathlig@next@tok##1%
   \mathlig@let@cs\mathlig@next@cmd{mathlig@look@#1#2##1}\let\mathlig@next\mathlig@gobble
  \fi 
  \ifx\mathlig@delim##2\mathlig@delim\relax\else
   \csname mathlig@chck@#1#2\endcsname##2\mathlig@delim
  \fi
 }%
 \ifx\mathlig@delim#2\mathlig@delim\else
  \mathlig@defcs{mathlig@back@#1#2}{\csname mathlig@back@#1\endcsname #2}%
 \fi
}%

\catcode`@\mathlig@atcode